\documentclass{article}
\usepackage{}
\usepackage{amssymb}
\usepackage[top=12mm]{geometry}
\usepackage{amsmath,amsthm,amssymb}
\usepackage{fancyhdr}
\usepackage{graphicx}
\usepackage{hyperref}
\usepackage{lipsum}
\usepackage{subfig}
\usepackage{varwidth}
\usepackage{enumitem}
\usepackage{mdwlist}%
\usepackage[noadjust]{cite}%
\usepackage{array}
\usepackage{epstopdf}
\usepackage{tabularx}
\usepackage{caption}
\usepackage{color, soul}
\usepackage{ulem}
\usepackage{cancel}
\usepackage{textcomp}
\usepackage{float}
\title{ Substrate Induced Optical Anisotropy in Monolayer MoS$_2$}

\author{
       \small{Wanfu Shen$^{1,2,3}$, Yaxu Wei$^{1,2,3}$, Chunguang Hu$^{2,3}$, C.B. L\'opez-Posadas$^{1}$, Michael Hohage$^{1}$, Lidong Sun$^{1*}$}
        \\ \scriptsize \noindent\textit{1. Institute of Experimental Physics, Johannes Kepler University Linz, A-4040 Linz, Austria}
	\\ \scriptsize \noindent\textit {2. State Key Laboratory of Precision Measuring Technology and Instruments, Tianjin University, Weijin Road 92,}
	\\ \scriptsize \noindent\textit {Nankai District, CN-300072 Tianjin, China}
	\\ \scriptsize \noindent\textit {3. Nanchang Institute for Microtechnology of Tianjin University, Weijin Road 92, Nankai District, 300072 Tianjin, China}
	\\ \scriptsize\textit{* Corresponding author: lidong.sun@jku.at}}
\date{}
\begin{document}

	\maketitle
	\newcommand{\upcite}[1]{\textsuperscript{\textsuperscript{\cite{#1}}}}  %\upcite
	%This is some preamble text that you enter
	%yourself.\footnote{First footnote.}\footnote{Second footnote.}
	\newcommand\hcancel[2][red]{\setbox0=\hbox{$#2$}%
	\rlap{\raisebox{.45\ht0}{\textcolor{#1}{\rule{\wd0}{1pt}}}}#2}
	\newcommand{\tabincell}[2]{\begin{tabular}{@{}#1@{}}#2\end{tabular}}
	\linespread{1.5}
    \renewcommand\thesection{\arabic{section}}%Roman

\section *{Abstract}	
In-plane optical anisotropy has been detected from monolayer MoS$_2$ grown on a-plane $(11\overline{2}0)$ sapphire substrate in the ultraviolet-visible wavelength range. Based on the measured optical anisotropy, the energy differences between the optical transitions polarized along the ordinary and extraordinary directions of the underlying sapphire substrate have been determined. The results corroborate comprehensively with the dielectric environment induced modification on the electronic band structure and exciton binding energy of monolayer MoS$_2$ predicted recently by first principle calculations. The output of this study proposes the symmetry as a new degree of freedom for dielectric engineering of the two-dimensional materials.
\vspace{10pt}

\section{Introduction}

Among the most studied two-dimensional (2D) semiconductors,
monolayer transition metal dichalcogenides (TMDCs) serve as the platform for fundamental studies in nanoscale and promise a wide range of potential applications.~\cite{mak2010Atomically, splendiani2010Emerging, Radisavljevic2011Single, Wang2012Electronics, Geim2013Van, qiu2013Optical, ugeda2014giant} Recently, the dielectric environment induced modification on the excitonic structures of monolayer TMDCs becomes a topic of intensive research efforts,~\cite{komsa2012effects, chernikov2014exciton, stier2016probing, rosner2016two, qiu2016screening, raja2017coulomb, kirsten2017band, cho2018environmentally, wang2018Colloquium} and the potential of the so called dielectric engineering in constructing novel optoelectronic devices has also been demonstrated.\cite{raja2017coulomb, utama2019dielectric, raja2019dielectric}

For freestanding monolayer TMDCs, due to quantum confinement and reduced dielectric screening, the Coulomb interactions between charge carriers are enhanced leading to a significant renormalization of the electronic structure and the formation of tightly bound excitons. While freestanding monolayer in vacuum representing the utmost reduction of dielectric screening, the electronic band structure and the binding energy between charge carriers in monolayer TMDCs can also be tuned by selecting dielectric environment. Indeed, first principle calculations predict a monotonic decrease of both electronic bandgap and exciton binding energy with increasing dielectric screening,\cite{komsa2012effects, qiu2013Optical, kirsten2017band, cho2018environmentally, wang2018Colloquium} which has also been observed experimentally\cite{stier2016probing, rosner2016two, raja2017coulomb, wang2018Colloquium}. Recently, by overlapping a homogeneous monolayer of MoS$_2$ (molybdenum disulfide) with the boundary connecting two substrates with different dielectric constants, an operational lateral heterojunction diode has been successfully constructed.\cite{utama2019dielectric} Even recently, a new concept named ``dielectric order'' has been introduced and its strong influence on the electronic transitions and exciton propagation has been illustrated using monolayer of WS$_2$ (tungsten disulfide)\cite{raja2019dielectric}. However, among these in-depth studies, the influence of the dielectric environment with a reduced symmetry has not been investigated,\cite{neupane2019plane} and its potential for realizing anisotropic modification on the electronic and optical properties of the monolayer TMDCs remains unexploited. In this letter, we report the breaking of the three-fold in-plane symmetry of the MoS$_2$ monolayer by depositing on the low-symmetry surface of sapphire, demonstrating the symmetry associated dielectric engineering of the 2D materials.

\section{Results and Discussions}

\begin{figure}[b]
\begin{center}
\includegraphics[width=6cm]{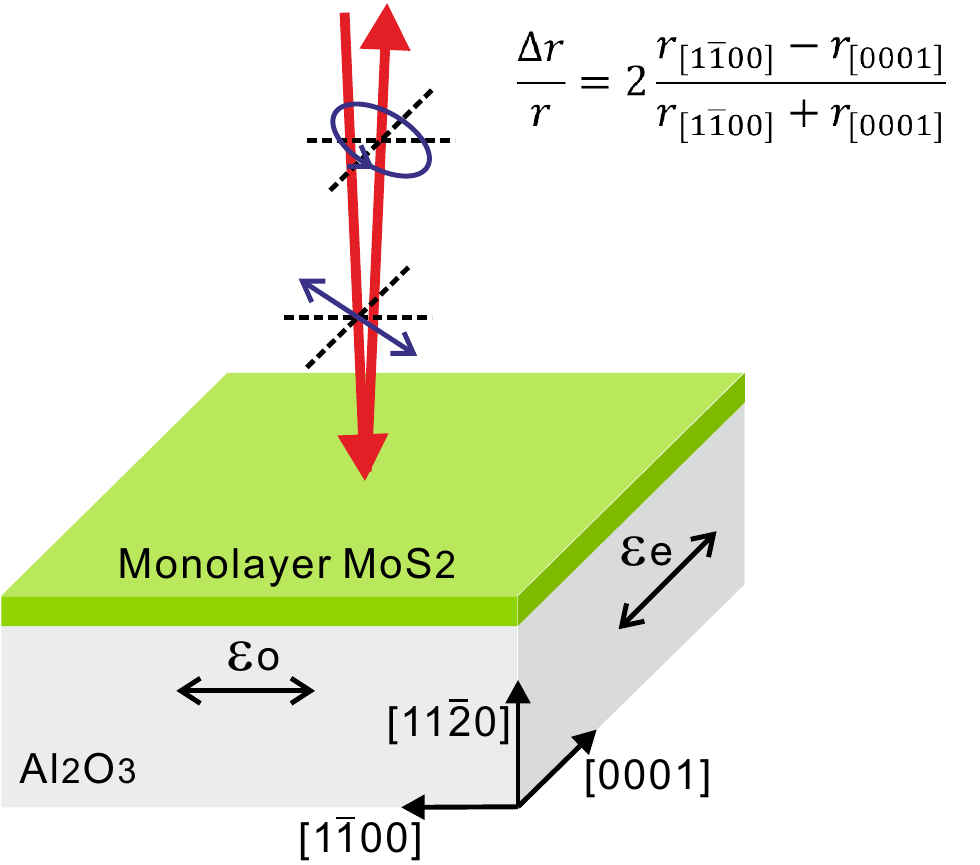}
\caption{(a)The setup of the RDS measurement and its alignment to the substrate.}
\label{figure1}
\end{center}
\end{figure}

Due to their attractive properties, sapphire crystals are widely applied in solid-state device fabrications and also among the substrate candidates for 2D semiconductors.\cite{singh2015al2o3, dumcenco2015large} Sapphire belongs to negative uniaxial crystals, i.e., its extraordinary dielectric function $\epsilon_e$ smaller than its ordinary dielectric function $\epsilon_o$.\cite{harman1994optical, yao1999anisotropic} So far, only c-plane (0001) sapphire substrate has been used to investigate its dielectric screening effects on the monolayer TMDCs.\cite{yu2015exciton,park2018direct} With isotropic in-plane dielectric properties defined by $\epsilon_o$, the underlying c-plane (0001) sapphire substrate induces a dielectric modification, which is laterally isotropic to monolayer TMDCs. In contrast, we prepared monolayer MoS$_2$ on a-plane ($11\overline{2}0$) sapphire substrate using chemical vapor deposition (CVD).\cite{supplemental} By selecting low symmetry a-plane sapphire as the substrate, we supply monolayer MoS$_2$ with an anisotropic dielectric environment defined by $\Delta \epsilon_{ext}=\epsilon_o-\epsilon_e$(see Fig.1). The resultant anisotropic modification was then investigate by measuring the optical anisotropy in the monolayer MoS$_2$ over the ultraviolet-visible (UV-Vis) range using reflectance difference spectroscopy (RDS),\cite{aspnes1985anisotropies,weightman2005reflection} which measures the reflectance difference between the light polarized along two orthogonal directions at close normal incidence (see Fig.1). This highly sensitive technology has been successfully applied to investigate the optical properties of ultra-narrow graphene nanoribbons.\cite{denk2014exciton} For the monolayer MoS$_2$ covered a-plane ($11\overline{2}0$) substrate, the RD signals can be described by the following equation:
\begin{equation}\label{Eq.1}
\frac{\Delta r}{r}=\frac{1}{2}\frac{r_{[1\overline{1}00]}-r_{[0001]}}{r_{[1\overline{1}00]}+r_{[0001]}}
\end{equation}
where $r_{[1\overline{1}00]}$ and $r_{[0001]}$ denote the reflectance of the light polarized along the $[1\overline{1}00]$ and the [0001] directions of the a-plane sapphire substrate, respectively.

\begin{figure}[h]
\begin{center}
\includegraphics[width=8cm]{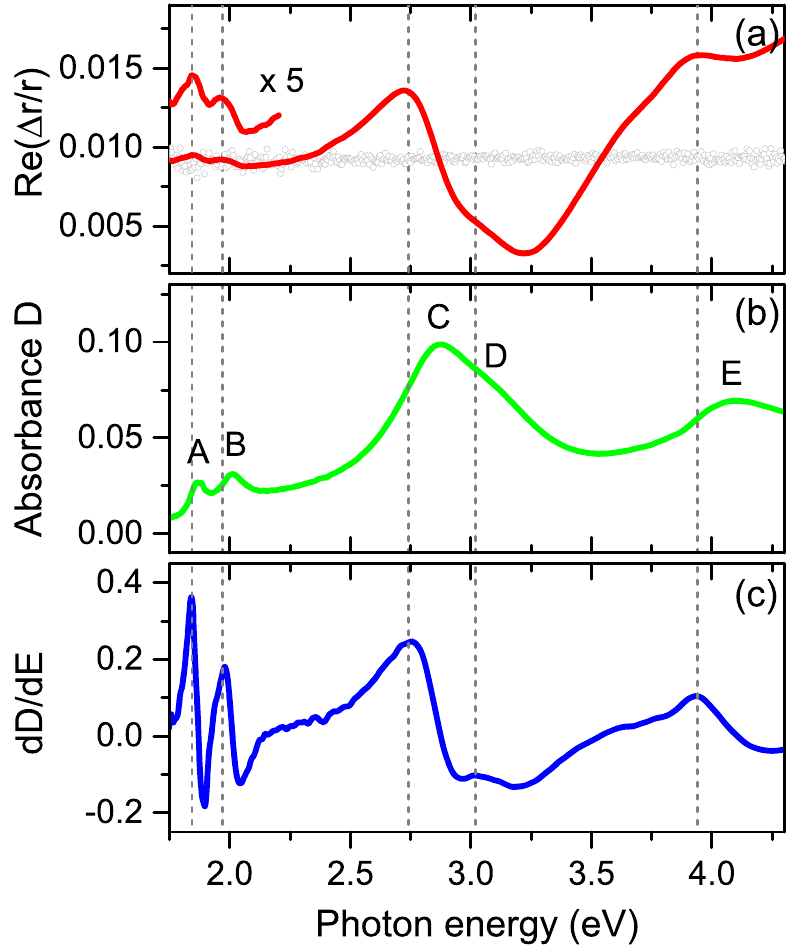}
\caption{(a) The RD spectrum taken from the monolayer MoS$_2$ on Al$_2$O$_3$($11\overline{2}0$), (b) The absorption spectrum (b) and its first derivative (c) measured from monolayer MoS$_2$ on Al$_2$O$_3$(0001).}
\label{figure2}
\end{center}
\end{figure}

After the systematic characterization using conventional techniques,\cite{supplemental} RDS measurement was then applied to investigate the optical anisotropy within the plane of the MoS$_2$ monolayer.The real part of the RD spectra measured from the bare Al2O3(11$\overline{2}$0) surface and the one covered by monolayer MoS$_2$ are plotted in Fig.~2(a), respectively. The bare Al2O3(11$\overline{2}$0) surface shows an optical anisotropy with an almost constant value which can be directly attributed to the in-plane birefringence of the a-plane sapphire substrate. Actually, the corresponding in-plane axis, namely [1$\overline{1}$00] and [0001] axis are parallel to the ordinary and extraordinary directions of sapphire, respectively. The result reveals thus the dielectric anisotropy $\Delta \epsilon_{ext}=\epsilon_o-\epsilon_e$ in a-plane sapphire substrate. Furthermore, additional optical anisotropy shows up from the a-plane sapphire substrate covered by monolayer MoS$_2$. It worth mentioning that, above the transparent sapphire substrate, the real part of the RD signal is predominantly associated with the anisotropy of the absorption of the monolayer MoS$_2$. For comparison, the absorption spectrum of the monolayer MoS$_2$ grown on a c-plane (0001) sapphire substrate\cite{supplemental} is plotted in Fig.~2(b). The spectrum exhibits typical absorption spectral line shape of monolayer MoS$_2$ with well resolved peaks indicated as A, B and C locating at 1.89\,eV, 2.03\,eV and 2.87\,eV, respectively. The peaks A and B are attributed to the electronic transitions from the spin-orbit split valence band (VB) to the conduction band (CB) around the critical points of K and K$'$ in the Brillouin zone, whereas the feature C is assigned to the transitions from VB to the CB in a localized region between critical points of K and $\Gamma$.\cite{qiu2013Optical, li2014measurement} Furthermore, two additional absorption features indicated with D and E rise at 3.06 and 4.09\,eV, respectively. A recent study combining experiments and first-principle calculation attributed the D and E features to the higher-lying interband transitions located at $\Gamma$ and K points of the Brillouin zone, respectively.\cite{Baokun2018Layer} Most importantly, the comparison between the spectra in Fig.~2(a) and (b) reveals apparent deviations of RD spectrum from the spectral line shape of absorption, indicating the observed anisotropy does not merely arise due to the polarization dependent reflectance of the substrate. In fact, the observed optical anisotropy should be the consequence of the anisotropic optical transitions of monolayer MoS$_2$. The result indicates thus the break of the pristine three-fold rotation symmetry of monolayer MoS$_2$. Indeed, the RD spectrum can be resembled precisely by the first derivative of the absorption spectrum (see Fig.~2(c)), regarding the overall line shape and, especially, the peak positions. Each peak in the RD spectrum coincides precisely with a local maximum on the first derivative curve, which is initiated by the rising slope of an absorption peak. Reminding the configuration
of RDS measurement in Fig.~1, this coincidence reveals that, at each critical point, the energy of optical transition for [1$\overline{1}$00]-polarized light (E$_{[1\overline{1}00]}$) is smaller than for [0001]-polarized light (E$_{[0001]}$), bringing in a positive energy shift $\Delta E = E_{[0001]}-E_{[1\overline{1}00]}$, accordingly.

In order to determine the energy shift $\Delta E$ for each individual optical transitions, the reflectance spectra of monolayer MoS$_2$ on a-plane (11$\overline{2}$0) sapphire substrate, namely $r_{[0001]}$ and $r_{[1\overline{1}00]}$, were simulated for the light polarized along [0001] and [1$\overline{1}$00] directions, respectively. For this purpose, a three-phase model composing vacuum, monolayer MoS$_2$, and a-plane (11$\overline{2}$0) sapphire substrate was used for the calculation (see Fig.~3(a)). The anisotropic dielectric functions of the a-plane sapphire substrate, namely $\epsilon_{[1\overline{1}00]}=\epsilon_o$ and $\epsilon_{[0001]}=\epsilon_e$, were measured using spectroscopic ellipsometry by Yao \emph{et al}.\cite{yao1999anisotropic} The dielectric function of the MoS$_2$ monolayer polarized along the [1$\overline{1}$00] direction of the sapphire substrate, i.e., $\epsilon_{\mathrm{MoS_2}[1\overline{1}00]}$, was deduced from the absorption spectrum of the one deposited on the isotropic Al$_2$O$_3$(0001) substrate (see Fig.~2(b)). To this end, the absorption spectrum was fitted by a superposition of multiple Lorentzian oscillators, each with well defined peak position $E_i$, amplitude $f_i$ and broadness $\Gamma_i$.\cite{li2014measurement} Subsequently, $\epsilon_{\mathrm{MoS_2}[0001]}$ polarized along the [0001] direction of the substrate is modeled by introducing a center energy shift $\Delta E_i$, an amplitude deviation $\Delta f_i$ and a line width difference $\Delta \Gamma_i$ for each individual Lorentzian oscillators constituting $\epsilon_{\mathrm{MoS_2}[1\overline{1}00]}$. The resultant reflectance spectra, namely $r_{[0001]}$ and $r_{[1\overline{1}00]}$, were subsequently used to calculate the corresponding RD spectrum using Eq.\,1. The Lorentzian parameters for each individual optical transitions were obtained by fitting the simulated RD spectrum with the one experimentally measured (see Fig.~3(b)). The real and imaginary parts of the dielectric function obtained for monolayer MoS$_2$ along the $[1\overline{1}00]$ and [0001] directions, respectively, are plotted in Fig.3(c). More details of the calculations can be found in Supplementary S3.

A close inspection at Fig.~3(b) confirms the systematic blue shift of $\epsilon_{\mathrm{MoS_2}[0001]}$ relative to $\epsilon_{\mathrm{MoS_2}[1\overline{1}00]}$. The energy shifts deduced for the A and B peaks are both around 0.02\,meV, and the $\Delta E$ increases to a value of $\sim$7.23\,meV and $\sim$15.88\,meV for the absorption features of C and E, respectively. The positive sign of the $\Delta E$s obtained agrees with the conclusion based on the coincidence between the RD and the differentiated absorption spectra. The weak feature D is excluded because its strong overlapping with the predominant broad C peak prevents deducing reliable $\Delta E$.

The observed optical anisotropy of monolayer MoS$_2$ can be explained by the anisotropic dielectric screening induced by the a-plane sapphire substrate. As introduced in the previous section, for the atomically thin semiconductors, it has been predicted that the surrounding dielectric environment modifies both their electronic band structure and their exciton binding energy significantly by the dielectric screening effect. However, near the band edge, the modification of the electronic band structure is largely compensated by the simultaneous alternation of the exciton states, resulting in only a moderate variation of the excitonic transition energy.\cite{cho2018environmentally} This effect, however, attenuates for optical transitions involving higher-lying bands, leading to a pronounced dielectric environment modification on the transition energy.\cite{cho2018environmentally, raja2019dielectric}

In the current case, being clipped between the air and substrate, the monolayer MoS$_2$ is exposed to the anisotropic dielectric environment invoked by the a-plane sapphire substrate, and its electronic band structure and exciton states become the objects of modification. For the a-plane sapphire substrate, at the static limit, the polarization dependent
dielectric constants read $\epsilon_{[1\overline{1}00]} = \epsilon_o = 3.064$ and $\epsilon_{[0001]}=\epsilon_e=3.038$. \cite{harman1994optical, yao1999anisotropic} The anisotropic dielectric environment indicated by $\Delta\epsilon=\epsilon_{1\overline{1}00}-\epsilon_{[0001]}=0.026$ may introduce the energy shifts $\Delta E$s between the [0001]- and $[1\overline{1}00]$-polarized optical transitions, and concomitantly the optical anisotropy of monolayer MoS$_2$ on a-plane sapphire substrate. Actually, the observed correlation between the $\Delta E$ and the $\Delta\epsilon$ agrees nicely with the previous results in the following respects: (1) Optical transition energy is predicted to decrease with increasing the dielectric permittivity. For a-plane sapphire substrate, $\epsilon_{[1\overline{1}00]}$is larger than $\epsilon_{[0001]}$. Consequently, the positive sign of $\Delta E=E_{[0001]}-E_{[1\overline{1}00]}$ obtained from each individual
features of RD spectrum is consistent with the prediction. (2) The environment induced modification on the optical transition energy is enhanced for the higher-lying interband transitions. In the experimental results presented here, the $\Delta E$ increases with optical transition energy dramatically. Actually, the $\Delta E$ associated with the higher-lying interband transitions (C and E) are several orders of magnitude larger than the ones related to the excitonic transitions below the bandgap (A and B).

\begin{figure}[h]
\begin{center}
\includegraphics[width=14cm]{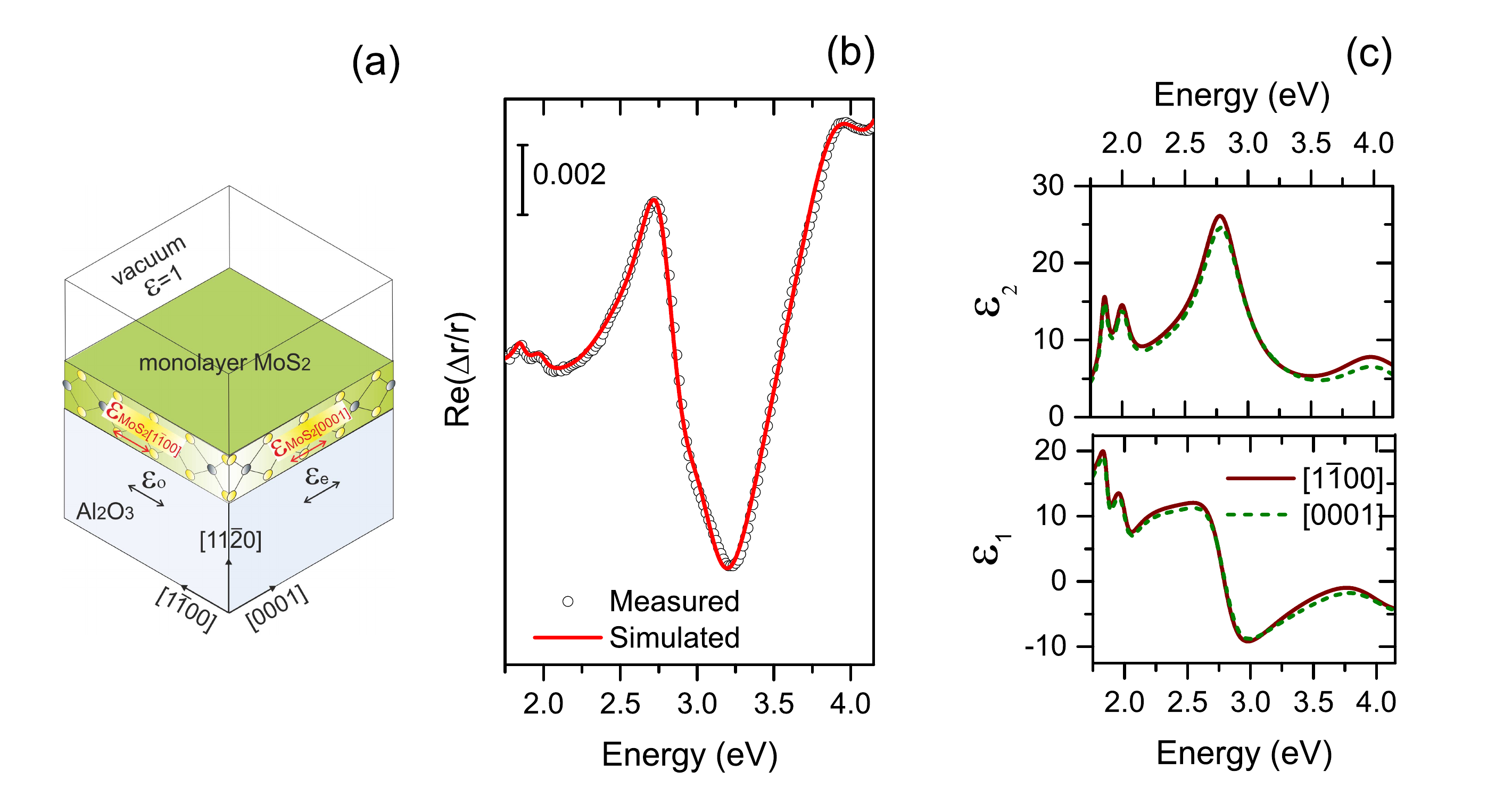}
\caption{(a) The schematic illustration of the three phase model, (b) the fitting between the simulated and the measured RD spectra, (c) the deduced dielectric functions of monolayer MoS$_2$ for light polarized along the [0001] and [1$\overline{1}$00] directions of a-plane sapphire substrate, respectively.}
\label{figure3}
\end{center}
\end{figure}

The RDS measurements have also been performed when the sample was enclosed in a vacuum chamber with a base pressure of $1 \times 10^{-9}$ mbar. Fig.~3 shows the RD spectra of the same sample of monolayer MoS$_2$ on $\mathrm{Al_2O_3} \ (11\overline{2}0)$ but measured in the atmosphere and vacuum, respectively. In comparison with the result obtained in air, the RD spectrum measured in vacuum shows clearly three new features between the B and C peaks (Fig.~3(a)). Based on their energetic positions, two of them can be attributed to the 2s and 3s states in the B exciton Rydberg series (see the indication in Fig.~4(a)).\cite{hill2015observation} The third feature, which appears as a shoulder at the right side of the peak B, is most probably associated with the 2s state of A exciton. This argument is supported by the observation that the energy interval between this feature and the 2s state of peak B is similar to the one between the 1s states of A and B exciton.\cite{hill2015observation} Besides, an intensification of RD signal can be recognized over the whole spectral range (Fig.~3(b)). The vacuum induced enhancement of the optical anisotropy can be explained by the improvement of the ``dielectric ordering''.\cite{raja2019dielectric,cadiz2017excitonic} In fact, the atmosphere may introduce dielectric disorder in following ways: (1) Adsorption of air molecules, such as water, on the top surface of monolayer MoS$_2$. (2) Molecular intercalation between the monolayer MoS$_2$ and the $\mathrm{Al_2O_3 \ (11\overline{2}0)}$ surface. \cite{he2012scanning} These processes may introduce a nonhomogeneous dielectric environment and weaken the regular anisotropic dielectric screening induced by the substrate. By reducing the ambient pressure in a vacuum, both adsorption and intercalation are hinted, leading to an improved dielectric ordering and enhanced anisotropic dielectric screening from the substrate. Consequently, the current experimental observation suggests also the potential of the substrate-induced optical anisotropy as a sensitive probe to the molecular adsorption and intercalation.

\begin{figure}[h]
\begin{center}
\includegraphics[width=10cm]{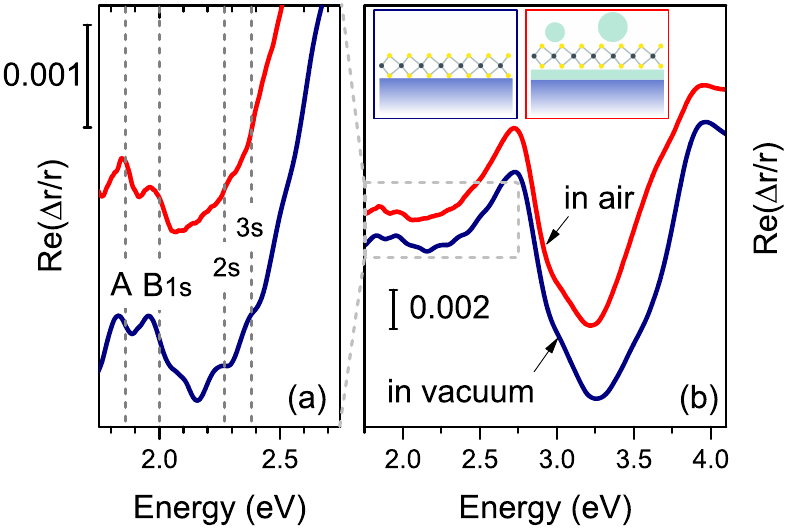}
\caption{(a) The RD spectra of monolayer MoS$_2$ on $\mathrm{Al_2O_3}(11\overline{2}0)$ measured in atmosphere and in vacuum, respectively, plotted in a narrow (a) and an extended (b) energy range. The interfacial structures without and with molecular adsorption and intercalation are schematically illustrated in the inset of (b).}
\label{figure4}
\end{center}
\end{figure}

\section{Conclusion}

In summary, optical anisotropy has been detected in monolayer MoS$_2$, which was deposited on a-plane $(11\overline{2}0)$ sapphire substrate by CVD. The revealed breaking of the intrinsic three-fold rotation symmetry of monolayer MoS$_2$ is associated with the anisotropic dielectric environment supplied by the underlying a-plane sapphire substrate. The resultant anisotropic modification of monolayer
MoS$_2$ has been quantitatively evaluated by determining the energy difference between optical transitions polarized along extraordinary and ordinary directions of the sapphire substrate. The general tendency based on the optical transitions over the ultraviolet-visible wavelength range is in good agreement with the first principle predication regarding the modification of the electronic band structure and exciton binding energy. Furthermore, the detailed optical anisotropy of monolayer MoS$_2$ shows a dependence on the ambient pressure, indicating its sensitivity to the molecular adsorption and intercalation. Although only one combination, namely monolayer MoS$_2$ and a-plane (11$\overline{2}$0) sapphire substrate has been investigated, the anisotropic dielectric modification should be a general phenomenon for atomically thin materials adjacent substrates with low symmetry. In addition to the magnitude and order, the symmetry of the dielectric environment may supply a new degree of freedom for dielectric engineering of the two-dimensional materials.

\section*{acknowledgements}
We acknowledge the financial support for this work by the Austrian Science Fund (FWF) with project number: P25377-N20. W.F.S. and Y.X.W. acknowledge the financial support of the China Scholarship Council (CSC), Y.X.W. acknowledges the financial support from Eurasia Pacific Uninet, C.B.L.P. acknowledges financial support from Consejo Nacional de cienciay Tecnologa, though the Becas Mixtas program.

\scriptsize%\small
\bibliographystyle{unsrt}
\bibliography{RDSofMoS2}

\end{document}

\maketitle
	\newcommand{\upcite}[1]{\textsuperscript{\textsuperscript{\cite{#1}}}}  %\upcite
	%This is some preamble text that you enter
	%yourself.\footnote{First footnote.}\footnote{Second footnote.}
	\newcommand\hcancel[2][red]{\setbox0=\hbox{$#2$}%
	\rlap{\raisebox{.45\ht0}{\textcolor{#1}{\rule{\wd0}{1pt}}}}#2}
	\newcommand{\tabincell}[2]{\begin{tabular}{@{}#1@{}}#2\end{tabular}}
	\linespread{1.5}
    \renewcommand\thesection{\arabic{section}}%Roman

\section{SAMPLE PREPARATIONS}
Monolayer MoS$_2$ has been synthesized on one side epi-ready a-plane (1120) and double side epi-ready c-plane (0001)sapphire substrates, respectively using low pressure chemical vapor deposition (CVD)\cite{lopez2019direct}. Systematic characterization were carried out using conventional techniques to confirm the monolayer thickness and the quality of the samples.

\section{SAMPLE CHARACTERIZATIONS}

The morphology of the monolayer MoS$_2$ grown on a-plane sapphire has been studied using ZEISS Leo Supra 35 scanning electron microscope (SEM) and the MoS$_2$ grown on c-plane sapphire was investigated using atomic force microscopy (Veeco Dimensions S3100) in tapping mode with a soft cantilever (TipsNano). All the the Raman and PL spectra were collected using a JY Horiba LabRAM Aramis VIS microscope with an excitation wavelength of 532 nm. These measurements were carried out in a confocal configuration using a $\times$100 objective lens and a 2400 grooves mm$^{-1}$ grating. For the measurement of differential reflectance spectroscopy (DRS), an inverted microscope (Nikon Eclipse Ti) was applied. To this end, a white light beam from a tungsten halogen lamp was reflected into a $\times$50 ultra-long working distance objective using a neutral density, achromatic beam splitter, and focused subsequently onto the sample surface. The reflected signals from the monolayer MoS2 covered areas (R$_{\mathrm{MoS2}}$ ) and the bare substrate areas (R$_{\mathrm{substrate}}$) were collected successively by the same objective and sent through the beam splitter back into the spectrometer. The DR spectra were then calculated using $\mathrm{DRS} = (R_{\mathrm{MoS2}}-R_{\mathrm{substrate}})/R_{\mathrm{substrate}}$.

\begin{figure}[h]
\begin{center}
\includegraphics[width=12cm]{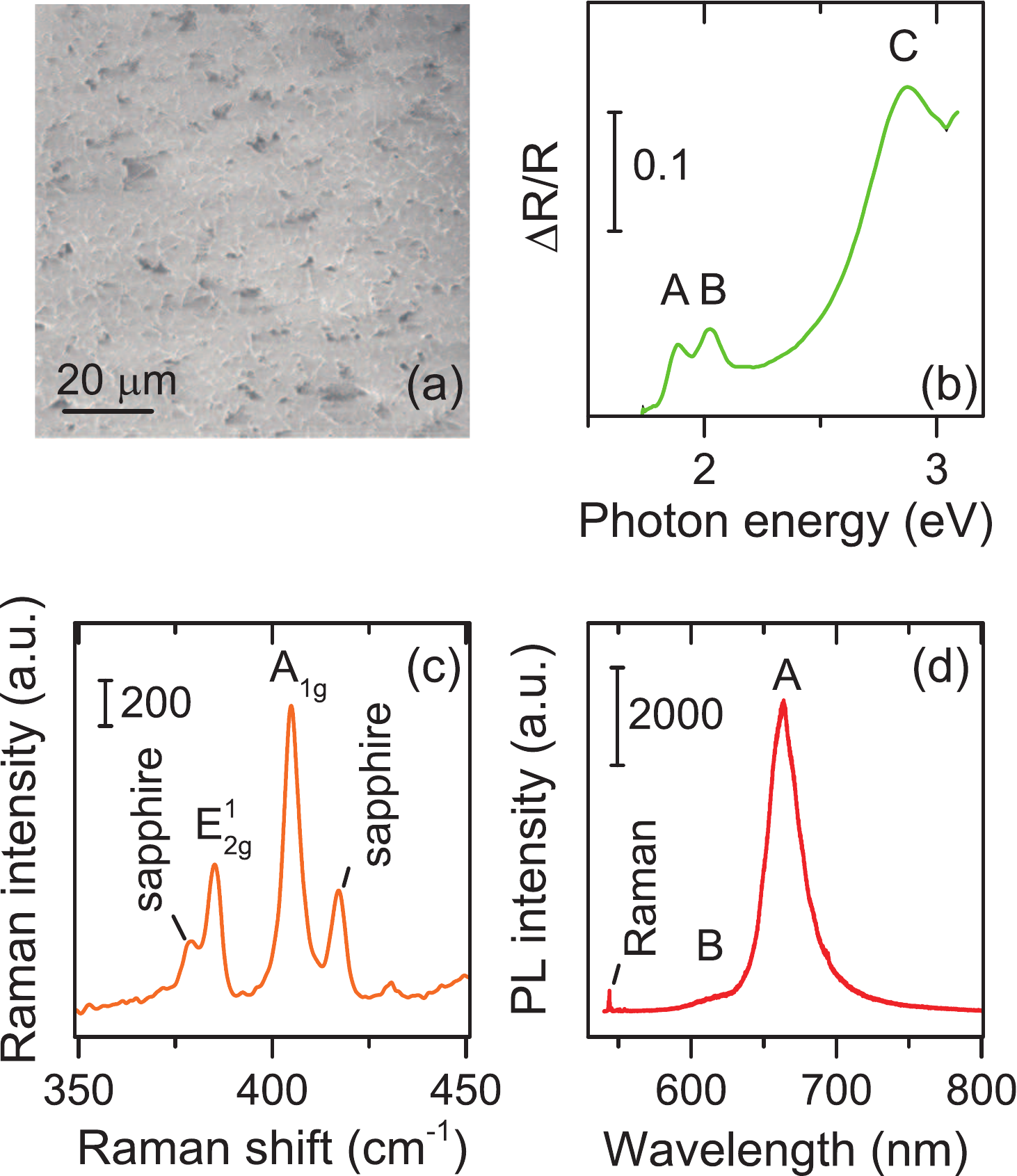}
\caption{The SEM image (a), DR spectrum (b), Raman Spectrum (c) and the PL spectrum (d) taken from the monolayer MoS$_2$ on the Al$_2$O$_3$(11$\overline{2}$0) surface.}
\label{figureS1}
\end{center}
\end{figure}

FigS.\ref{figureS1} summaries the results obtained from conventional characterization on the monolayer MoS$_2$ grown on Al2O3(11$\overline{2}$0). The SEM image presented in FigS.\ref{figureS1}(a) shows the surface is fully covered by the MoS2 domains with size around several \textmu m. Due to the transparency of the sapphire substrate, the corresponding DR spectrum in FigS.\ref{figureS1}(b) exhibits typical absorption spectral line shape of MoS$_2$ monolayer with well resolved peaks indicated as A, B and C locating at 1.89 eV, 2.03 eV and 2.87 eV, respectively. The peaks A and B are attributed to the electronic transitions from the spin-orbit split valence band (VB) to the conduction band (CB) around the critical points of K and K$'$ in the Brillouin zone, whereas the feature C is assigned to the transitions from VB to the CB in a localized region between critical points of K and $\Gamma$\cite{li2014measurement}. The Raman spectrum plotted in FigS.\ref{figureS1}(c) exhibits the characteristic E$^1_{2g}$ and A$_{1g}$ peaks at 385.6 and 404.4 cm$^{-1}$, respectively. The interval between these two vibration modes is $\sim$19 cm$^{-1}$ confirming the monolayer thickness of the MoS$_2$ film \cite{tonndorf2013photoluminescence}. This conclusion is further verified by the strong characteristic PL emission presented in FigS.\ref{figureS1}(d).

\begin{figure}[h]
\begin{center}
\includegraphics[width=12cm]{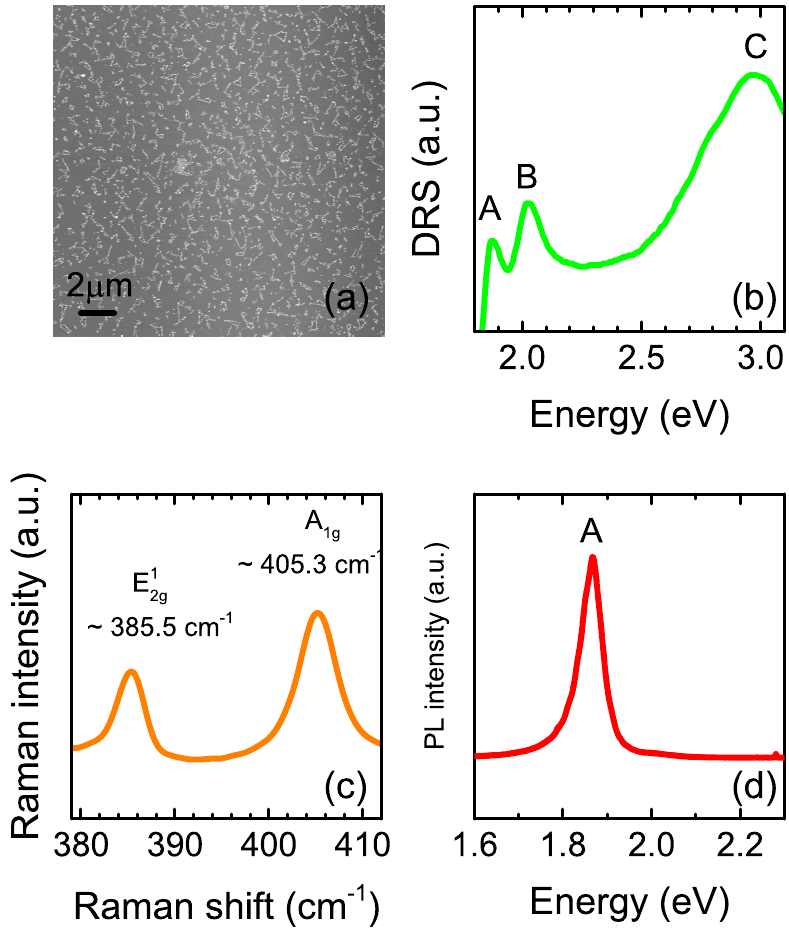}
\caption{(a)The AFM image (a), DR spectrum (b), Raman spectrum (c), and PL spectrum (d) taken from the monolayer MoS$_2$ on the Al$_2$O$_3$ (0001) surface.}
\label{figureS2}
\end{center}
\end{figure}

FigS.\ref{figureS2} shows the characterization results carried out on the monolayer MoS$_2$ grown on the Al$_2$O$_3$(0001) surface, confirming the formation of homogeneous monolayer MoS$_2$ with typical optical and vibrational properties.

\section{CALCULATIONS}

\begin{figure}[h]
\begin{center}
\includegraphics[width=12cm]{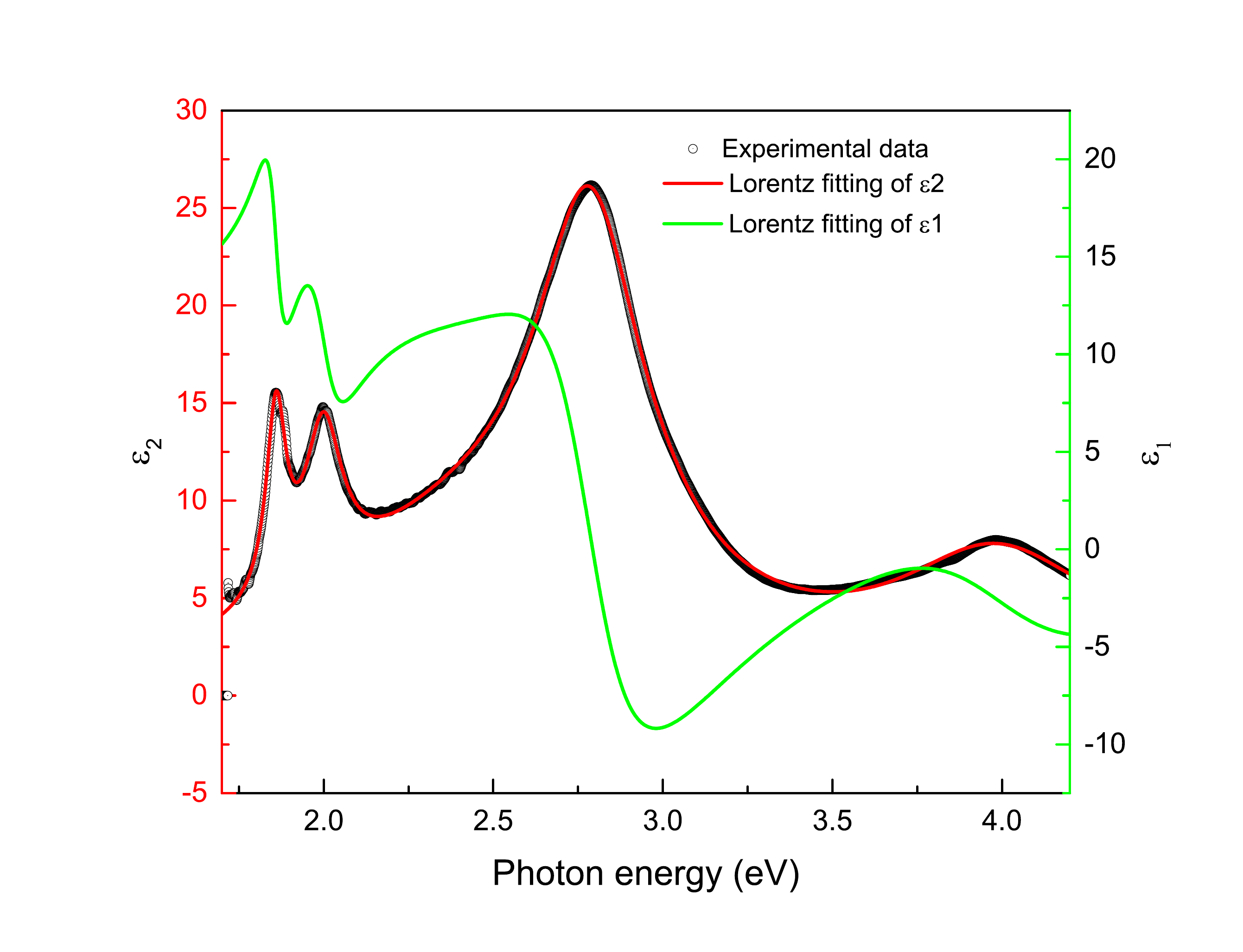}
\caption{(a)The AFM image (a), DR spectrum (b), Raman spectrum (c), and PL spectrum (d) taken from the monolayer MoS$_2$ on the Al$_2$O$_3$ (0001) surface.}
\label{figureS3}
\end{center}
\end{figure}

Firstly, we measured monolayer MoS$_2$ grown on c-plane (0001) sapphire with same conditions of the one on a-plane (11$\overline{2}0$) sapphire using ellipsometry (J.A.Woollam, M2000) to obtain its dielectric function, which was used as polarization-resolved complex dielectric function $\epsilon_{\mathrm{MoS_2}[1\overline{1}00]}$ along $[1\overline{1}00]$ direction. The experimental dielectric function as shown by the black circles in FigS.\ref{figureS3}. Secondly, we used multiple lorentz oscillators to reproduce the measured $\epsilon_{\mathrm{MoS_2}[1\overline{1}00]}$ by
\begin{equation}\label{lortenz}
\begin{split}
\epsilon^{Lor}_{2,i}(E)=\dfrac{f_i\Gamma_iE}{(E^2_i-E^2)^2+\Gamma^2_iE^2},\\
\epsilon^{Lor}_{1,i}(E)=\dfrac{f_i(E^2_i-E^2)}{(E^2_i-E^2)^2+\Gamma^2_iE^2},
\end{split}
\end{equation}
where $E_i$ is the central energy, $f_i$ is the strength, and $\Gamma_i$ is the line width of the $i$th Lorentz oscillators. Using least-square method, the fitting results as shown by the solid lines in FigS.\ref{figureS3}. The detailed parameters of the Lorentz oscillators were listed in TabS.1.

\begin{table}[!ht]
   \centering
    \caption{The parameters of fitted Lorentz oscillators}
    \label{lorentz parameters}
 \begin{tabular}{cccc}
  \toprule
  $i$ & Central Energy ($E_i$) & Strength ($f_i$) & Line width ($\Gamma_i$)\\
  \midrule
  1 & 1.86 &  1.2719  & 0.0713  \\
  2 & 2.00 &  2.1944  & 0.13483  \\
  3 & 2.50 &  20.213   & 1.1228    \\
  4 & 2.79 &  21.604 & 0.3879    \\
  5 & 3.00 &  0.61728  & 0.3918    \\
  6 & 3.18 &  $1.27 \times 10^{-9} $ & 16.56    \\
  7 & 4.00 & 17.756   & 0.73109    \\
  8 & 4.60 & 5.4413   & 0.6099   \\
  \bottomrule
  \end{tabular}
 \end{table}

~\\
~\\
~\\

Thus we calculated the reflectance along the two crystalline axes of sapphire by

\begin{equation}\label{lortenz}
\begin{split}
r_{01\_x}=\dfrac{n_0-n_{\mathrm{MoS_2}\_x}}{n_0+n_{\mathrm{MoS_2}\_x}},\\
r_{12\_x}=\dfrac{n_{\mathrm{MoS_2}\_x}-n_o}{n_{\mathrm{MoS_2}\_x}+n_o}, \\
r_{01\_y}=\dfrac{n_0-n_{\mathrm{MoS_2}\_y}}{n_0+n_{\mathrm{MoS_2}\_y}}, \\
r_{12\_y}=\dfrac{n_{\mathrm{MoS_2}\_y}-n_e}{n_{\mathrm{MoS_2}\_y}+n_e}, \\
\end{split}
\end{equation}
where $n_{\mathrm{MoS_2}\_x}$ indicates the refraction indexes of MoS$_2$ along the [1$\overline{1}00$] directions of Al$_2$O$_3$ (11$\overline{2}$0), with refractive index of $\epsilon_o$. $n_{\mathrm{MoS_2}\_y}$ indicates the refraction indexes of MoS$_2$ along the [0001] directions, with refractive index of $\epsilon_e$. Similarly, we modeled the dielectric function of MoS$_2$ along [0001] direction by adding shifting parameters on each Lorentz oscillators formulated long [1$\overline{1}00$] direction:
\begin{equation}\label{lortenz shift}
\begin{split}
\epsilon^{Lor}_{2,i}(E)=\dfrac{(f_i+\Delta f_i)(\Gamma_i+\Delta \Gamma_i)E}{((E_i+\Delta E_i)^2-E^2)^2+(\Gamma_i+\Delta\Gamma_i)^2E^2},\\
\epsilon^{Lor}_{1,i}(E)=\dfrac{(f_i+\Delta f_i)((E_i+\Delta E_i)^2-E^2)}{((E_i+\Delta E_i)^2-E^2)^2+(\Gamma_i+\Delta\Gamma_i)^2E^2},
\end{split}
\end{equation}
 where the $E_i$, $f_i$, and $\Gamma_i$ are as same as the values of modeling dielectric function along [1$\overline{1}00$] directions as listed in Tab. 1S.  By substituting Eq.\ref{lortenz}-eq.\ref{lortenz shift} into the definition of RDS in the main text of  Eq. 1, we will obtain
 \begin{equation}\label{fitting}
 \mathrm{Re}(\dfrac{\Delta r}{r})=F(\Delta E_1, \Delta f_1, \Delta \Gamma_1,...,\Delta E_i, \Delta f_i, \Delta \Gamma_i, \Delta E_8, \Delta f_8, \Delta \Gamma_8 ).
 \end{equation}
We fitted Eq. \ref{fitting} using least-square method and the fitted results was shown in Fig.3(b) in the main text. The fitting parameters was listed in Tab.2S.

\begin{table}[t]
   \centering
    \caption{The fitted shifting parameters}
    \label{fitting lorentz parameters}
 \begin{tabular}{ccccc}
  \toprule
  $i$ & Central Energy ($E_i$) & Central Energy Shift ($\Delta E_i$ (meV)) & Strength Shift($\Delta f_i$) & Line width Shift ($\Delta\Gamma_i$)\\
  \midrule
  1 & 1.86 &  0.01728  & -0.03705 & $-1.7913\times 10^{-4}$ \\
  2& 2.00 &  0.0169  & -0.05543 & $-1.4248\times 10^{-4}$ \\
  3& 2.50 &  -25.74  & -1.4438 & -0.007 \\
  4& 2.79 &  7.23  & 0.18674 & 0.01538 \\
  5& 3.00 &  158  & 1.5585 & 0.2462 \\
  6& 3.18 &  -91.76  & -8.1327 & -9.1962 \\
  7& 4.00 &  15.88  & -2.2009 & -0.01101 \\
  8& 4.6 &  74.44  & -3.6615 & -0.40894 \\
  \bottomrule
  \end{tabular}
 \end{table}

\scriptsize%\small
\bibliographystyle{unsrt}
\bibliography{RDSofMoS2-SM}